\documentclass[preprint,10pt]{elsarticle}
\usepackage[margin=1.5in]{geometry}
\usepackage{amsmath,amssymb,amsthm,pifont,verbatim}
\usepackage{graphicx,subfigure}
\usepackage{epic,fancybox}
\usepackage{epsfig}
\usepackage{epstopdf}
\usepackage{url}
\usepackage{color}
\usepackage{algpseudocode}
\usepackage{algorithm}
\usepackage{scrextend}
\usepackage{hyperref}
\usepackage{tabu}
\usepackage{appendix}
\usepackage{lineno}

\journal{Journal of Computational Physics}

\begin{document}

\begin{frontmatter}

\title{Efficient traveltime solution of the acoustic TI eikonal equation}

\author{Umair bin Waheed\corref{cor1}}
\ead{umairbin.waheed@kaust.edu.sa}
\cortext[cor1]{Corresponding author}

\author{Tariq Alkhalifah}
\ead{tariq.alkhalifah@kaust.edu.sa}

\author{Hui Wang}
\ead{hui.wang@kaust.edu.sa}

\address{King Abdullah University of Science and Technology, 4700 KAUST, 23955-6900, Thuwal, Saudi Arabia}

\begin{abstract}
Numerical solutions of the eikonal (Hamilton-Jacobi) equation for transversely isotropic (TI) media are essential for imaging and traveltime tomography applications. Such solutions, however, suffer from the inherent higher-order nonlinearity of the TI eikonal equation, which requires solving a quartic polynomial for every grid point. Analytical solutions of the quartic polynomial yield numerically unstable formulations. Thus, we need to utilize a numerical root finding algorithm, adding significantly to the computational load. Using perturbation theory we approximate, in a  first order discretized form, the TI eikonal equation with a series of simpler equations for the coefficients of a polynomial expansion of the eikonal solution, in terms of the anellipticity anisotropic parameter. Such perturbation, applied to the discretized form of the eikonal equation, does not impose any restrictions on the complexity of the perturbed parameter field. Therefore, it provides accurate traveltime solutions even for models with complex distribution of velocity and anisotropic anellipticity parameter, such as that for the complicated Marmousi model. The formulation allows for large cost reduction compared to using the exact TI eikonal solver. Furthermore, comparative tests with previously developed approximations illustrate remarkable gain in accuracy in the proposed algorithm, without any addition to the computational cost.
\end{abstract}


\end{frontmatter}

\section{Introduction}
\label{sec:1}
Eikonal equations arise in many practical applications ranging from classical mechanics to optimal control. These include problems in geometrical optics, computer vision, obstacle navigation, manufacturing of computer chips, image processing, etc.~\cite{sethian_level_1999}. In seismology, solutions of the eikonal equations are routinely employed to compute traveltimes for numerical modeling and migration of seismic waves.
For example, traveltime tables are needed for Kirchhoff modeling and migration algorithms. It is also vital to many velocity estimation applications, such as reflection tomography.

Eikonal equation is a non-linear partial differential equation (PDE) obtained from the first term of the Wentzel-Kramers-Brillouin expansion of the wave equation. It represents a class of the Hamilton-Jacobi (HJ) equations. Among the most widely used methods for solving this equation is ray tracing~\cite{cerveny_seismic_2005} and finite-difference approximations of the eikonal itself. Finite-difference eikonal solvers provide a robust and relatively fast method of traveltime computation compared to ray tracing~\cite{sava_3-d_2001}. They also avoid the computationally expensive step of traveltime interpolation to a regular grid which imaging applications require. Therefore, a lot attention has been paid over the years to compute first arrival traveltimes by numerically solving the eikonal equation~\cite{vidale_finite-difference_1990,van_trier_upwind_1991,sethian_3-d_1999,zhao_fast_2005}. Although the computed solutions are limited to first arrival traveltimes (the viscosity solution), these eikonal solvers can be extended in several different ways to image multiple arrivals~\cite{bevc_imaging_1997}.

Since sedimentary rocks cause anisotropic wave propagation~\cite{crampin_review_1981}, accurate seismic imaging requires that traveltime computation honors anisotropy whenever it significantly affects the kinematics of the wave propagation~\cite{qian_paraxial_2001}. This anisotropic behavior of waves is related to thin layers of isotropic or transversely isotropic (TI) rocks of different properties with respect to the dominant wavelength of our waves. Due to the gravity of the Earth, the layers are naturally aligned horizontally which gives rise to a TI medium with vertical axis of symmetry (VTI). However, tectonic forces inside the crust 
as well as migration of salt bodies may cause these layers to tilt and rotate, resulting in a TI medium with tilted axis of symmetry (TTI). The symmetry direction in this case is set normal to the layering~\cite{alkhalifah_building_2000}. Such a model represents one of the most effective approximations to the Earth subsurface.

Among the different proposed approaches for obtaining numerical solution of the eikonal equation, the most utilized are the fast marching method~\cite{sethian_level_1999} and the fast sweeping method~\cite{zhao_fast_2005}. These methods were originally proposed to solve the isotropic eikonal equation. However, there have been modifications proposed to these methods aiming to solve the anisotropic eikonal equations~\cite{sethian_ordered_2003,tsai_fast_2003,konukoglu_recursive_2007,qian_fast_2007}. These methods, at most, attempt to solve the tilted elliptically anisotropic (TEA) eikonal equation, which is a degenerate case of the TI eikonal equation.

Computing a stable solution for the TI eikonal equation using finite-difference methods is a complicated task, because such a process requires finding the roots of a quartic polynomial for every grid point. It also requires selection of the correct outgoing P-wave solution of the quartic polynomial. Recently, methods have been developed to solve the TI eikonal equation using an iterative solver~\cite{ma_qp_2013,waheed_fast_2013}. Such a process focuses on the simpler TEA eikonal equation at the first iteration and uses it as a building block for computing solution to the more complicated TI eikonal equation.

The P-wave traveltime in TTI media depends on parameters in symmetry direction, the on-axis velocity $v_0$, the on-axis normal moveout (NMO) velocity $v_{nmo}=v_0 \sqrt{1+2\delta}$, where $\delta$ is the anisotropy parameter, and the anellipticity anisotropy parameter $\eta$. In addition, it also depends on the angle, $\theta$, that the symmetry axis makes with the vertical. For the 3D case, it also depends on the azimuthal angle, $\phi$, of the plane containing the symmetry axis with respect to the $x$-axis~\cite{tsvankin_moveout_1997}.

In this paper, we derive a traveltime approximation for TTI media based on the perturbation of the anellipticity parameter for the discretized form of the TI eikonal equation. This allows the flexibility of incorporating heterogeneous $\eta$ structures into the modeling scheme, unlike some of the earlier attempts~\cite{alkhalifah_scanning_2011,waheed_diffraction_2013} that dealt with smoothly varying $\eta$ fields. The convergence rate of the proposed algorithm can be accelerated by the use of Shanks transform~\cite{bender_advanced_1999}. This results in a fast and accurate algorithm for traveltime computations for TTI media. The numerical algorithm is based on the fast sweeping method which ensures stability and robustness.

The rest of the paper is organized as follows. In Section~\ref{sec:2}, we derive the traveltime approximation based on perturbation of the $\eta$ parameter. This is followed by a description of the algorithm in Section~\ref{sec:3}. Section~\ref{sec:4} is devoted to numerical tests of the proposed algorithm. We first consider a simple homogeneous TTI model that offers great insight into the accuracy and convergence properties of the algorithm. Then, we perform tests on more realistic models including the VTI Marmousi model~\cite{alkhalifah_anisotropic_1997} and the BP TTI model~\cite{shah_2007}.

\section{Theory} 
\label{sec:2}

\subsection{The TTI eikonal equation}

The 2D eikonal equation for VTI media, under the acoustic assumption, is given as~\cite{alkhalifah_acoustic_2000}:
\begin{equation}
v_{nmo}^2(1+2\eta)\biggl(\frac{\partial \tau}{\partial x}\biggr)^2 + v_0^2\biggl(\frac{\partial \tau}{\partial z}\biggr)^2 \left(1-2\eta v_{nmo}^2\biggl(\frac{\partial \tau}{\partial x}\biggr)^2\right)= 1,
\label{eq:vtieikonal}
\end{equation}
where $\tau(x,z)$ is the traveltime measured from source to a receiver point $(x,z)$, $v_0$ and $v_{nmo}$ are the vertical and NMO velocities measured along the axis of symmetry and $\eta$ is the anellipticity parameter.

Representing TI media using $v_0$, $v_{nmo}$, and $\eta$ allows us to simplify the description of the wave equation under the acoustic assumption. It also allows for direct understanding
of the influence of anisotropy on critical seismic applications like imaging and tomography. This is in consideration of the fact that seismic data are acquired along one surface, the Earth surface, and TI has a preferred direction, specifically normal to that surface. These parameters are defined with respect to the elastic coefficients as follows:
\begin{equation}
v_0 = \sqrt{\frac{c_{33}}{\rho}}, \hspace{4cm}
\end{equation}
\begin{equation}
v_{nmo} = \sqrt{\frac{c_{13}\left(c_{13} + 2c_{55}\right) + c_{33}c_{55}}{\left(c_{33} - c_{55}\right) \rho}}, \hspace{1.2cm}
\end{equation}
\begin{equation}
\eta = \frac{c_{11}\left(c_{33} - c_{55}\right)}{2 c_{13}\left(c_{13} + 2c_{55}\right)+2c_{33}c_{55}} - \frac{1}{2},
\end{equation}
where $\rho$ is the density, and $c_{13}, c_{33}, c_{55}$ denote the independent components in the stiffness tensor, represented under the Voigt notation. For more details on the relationship between the elastic coefficients and anisotropy parameters, refer to~\cite{thomsen_weak_1986}.

For TTI media, the traveltime derivatives in Equation~(\ref{eq:vtieikonal}) are rotated by the operator:
\begin{equation*}
\begin{bmatrix} 
  \cos\theta     & \sin\theta\\ 
  -\sin\theta & \cos\theta 
\end{bmatrix},
\end{equation*}
where $\theta$ is the tilt angle that the symmetry axis makes with the vertical. Then, the eikonal equation for 2D TTI media is expressed as:
\begin{equation}
\begin{aligned}
&{v_{nmo}^2} (1+2 \eta) \,{\left(\cos\theta \frac{\partial \tau}{\partial x} + \sin\theta \frac{\partial \tau}{\partial z}\right)^2 } + {{{v_0}}^2}\,{\left( \cos\theta \frac{\partial \tau}{\partial z}-\sin\theta \frac{\partial \tau}{\partial x} \right)^2}\,\\
& \times \left( 1 - 2 \eta {v_{nmo}^2} \,{ \left( \cos\theta \frac{\partial \tau}{\partial x} +\sin\theta \frac{\partial \tau}{\partial z} \right)^2} \right)=1.
\label{eq:ttieikonal}
\end{aligned}
\end{equation}
Here we consider the 2D case for simplicity of illustration. The methodology developed below can be easily extended to the eikonal equation for 3D TTI media.
 
The complication in obtaining numerical solution to Equation~(\ref{eq:ttieikonal}) arises due to the presence of terms involving spatial derivatives to the power of four. Under the finite-difference approximation, this requires solving a quartic polynomial for every grid node. This is followed by the need to choose the correct outgoing P-wave solution among the four roots of the quartic polynomial. These steps lead to a significant increase in computational load~\cite{waheed_efficient_2013}. 

In order to alleviate this problem, Stovas and Alkhalifah~\cite{stovas_new_2012} proposed the use of traveltime expansion as a function of $\eta$ by perturbing the TTI eikonal equation~(\ref{eq:ttieikonal}). This required solving a much simpler TEA eikonal equation for the zeroth order traveltime approximation and a bunch of linear PDEs for the higher order terms of the expansion. The downside was that the perturbation parameter $\eta$ was not allowed to vary spatially, hence only a smoothly varying $\eta$ field could be used in the modeling scheme.

We propose the application of perturbation expansion on the discretized form of Equation~(\ref{eq:ttieikonal}). The advantages of using this approach are two folds. First, it lowers the computational cost significantly, and second, it removes the smoothness constraint on $\eta$. The zeroth order term in the new expansion still requires solving a TEA eikonal equation. However, the computational speed up is obtained due to the derivation of analytical expressions for the higher order terms of the perturbation expansion.

\subsection{Traveltime approximation}

The first order finite-difference approximation of the traveltime derivatives with respect to the spatial variables $x$ and $z$ are given as:
\begin{equation}
\begin{aligned}
& \frac{\partial \tau}{\partial x} = \left(\frac{\tau_{i,j}-\tau_{x\,min}}{\Delta x}\right)s_x,\qquad
& \frac{\partial \tau}{\partial z} = \left(\frac{\tau_{i,j}-\tau_{z\,min}}{\Delta z}\right)s_z,
\label{eq:discretization}
\end{aligned}
\end{equation}
$$i = 2,3,...,I-1, \qquad j = 2,3,...,J-1.$$
In the above discretization, $\Delta x$ and $\Delta z$ denote grid spacing along the $x$ and $z$ directions, respectively, whereas $\tau_{i,j}$ denotes the sought traveltime solution at grid node $(i,j)$, while
\begin{equation}
\begin{aligned}
\tau_{x\,min} & = \mathrm{min}\left(\tau_{i-1,j},\tau_{i+1,j}\right), \qquad \tau_{z\,min} & = \mathrm{min}\left(\tau_{i,j-1},\tau_{i,j+1}\right),
\label{eq:neighbor}
\end{aligned}
\end{equation}
and
\begin{equation}
\begin{aligned}
s_x & = \begin{cases}
+1, & \text{if} \; \tau_{x\,min} = \tau_{i+1,j}\\
-1, & \text{if} \; \tau_{x\,min} = \tau_{i-1,j}
\end{cases},\\
s_z & = \begin{cases}
+1, & \text{if} \; \tau_{z\,min} = \tau_{i,j+1}\\
-1, & \text{if} \; \tau_{z\,min} = \tau_{i,j-1}
\end{cases}
.
\label{eq:sign}
\end{aligned}
\end{equation}

In Equation~(\ref{eq:neighbor}), $\tau_{i-1,j}$ and $\tau_{i+1,j}$ represent traveltimes at the neighboring grid nodes of the grid $(i,j)$ along the $x$ direction, while $\tau_{i,j-1}$ and $\tau_{i,j+1}$ denote traveltime values at the neighboring grid nodes of the grid point $(i,j)$ along the $z$ direction. The sign variables $s_x$ and $s_z$ ensure that an upwind discretization is used. The total number of grid points along the $x-$ and $z-$directions are denoted by $I$ and $J$, respectively.

Substituting the finite-difference discretization for traveltime derivatives from Equation~(\ref{eq:discretization}) into Equation~(\ref{eq:ttieikonal}), we get the discretized form of the TTI eikonal equation:
\begin{equation}
\begin{aligned}
& {v_{nmo}^2} (1+2 \eta_{i,j}) \,{\left(\cos\theta \left(\frac{\tau_{i,j}-\tau_{x\,min}}{\Delta x}\right)s_x + \sin\theta \left(\frac{\tau_{i,j}-\tau_{z\,min}}{\Delta z}\right)s_z\right)^2 }\\
& +  {{{v_0}}^2}\,{\left( \cos\theta \left(\frac{\tau_{i,j}-\tau_{z\,min}}{\Delta z}\right)s_z-\sin\theta \left(\frac{\tau_{i,j}-\tau_{x\,min}}{\Delta x}\right)s_x \right)^2}\\
& \times \left( 1 - 2 \eta_{i,j} {v_{nmo}^2} \,{ \left( \cos\theta \left(\frac{\tau_{i,j}-\tau_{x\,min}}{\Delta x}\right)s_x +\sin\theta \left(\frac{\tau_{i,j}-\tau_{z\,min}}{\Delta z}\right)s_z \right)^2} \right)=1.
\end{aligned}
\label{eq:discretetti}
\end{equation}
Note that in Equation~(\ref{eq:discretetti}) and what follows, $v_{0}$, $v_{nmo}$, and $\theta$ correspond to their values at grid point $(i,j)$. The subscript $(i,j)$ has been omitted from these variables to avoid complexity of notation.

In order to solve Equation~(\ref{eq:discretetti}) for traveltime $\tau_{i,j}$, we propose a trial solution based on the perturbation of $\eta$ parameter around the point $\eta_{i,j}=0$. This is expressed as:
\begin{equation}
\tau_{i,j} \approx \tau_{i,j}^{(0)} + \tau_{i,j}^{(1)}\, \eta_{i,j} + \tau_{i,j}^{(2)}\, \eta_{i,j}^2, \, \footnote{Note that the superscript inside parenthesis denotes the order of the term in the expansion (For ex. $\tau_{i,j}^{(1)}$ represents the first order term of the traveltime approximation), whereas $\eta_{i,j}^2 $denotes the fact that the variable $\eta_{i,j}$ is raised to power 2.}
\label{eq:trial}
\end{equation}
where $\tau_{i,j}^{(0)}, \tau_{i,j}^{(1)},$ and $\tau_{i,j}^{(2)}$ are the coefficients of expansion with dimensions of traveltime. We have considered only the first three terms of the expansion for ease of illustration.

Next, plugging the trial solution from Equation~(\ref{eq:trial}) into the discretized TTI eikonal equation~(\ref{eq:discretetti}), we get:
\begin{equation}
\resizebox{0.9\hsize}{!}{$
\begin{aligned}
& {v_{nmo}^2} (1+2 \eta_{i,j}) \,{\left(\cos\theta \left(\frac{\tau_{i,j}^{(0)} + \tau_{i,j}^{(1)} \eta_{i,j} + \tau_{i,j}^{(2)} \eta_{i,j}^2-\tau_{x\,min}}{\Delta x}\right)s_x + \sin\theta \left(\frac{\tau_{i,j}^{(0)} + \tau_{i,j}^{(1)} \eta_{i,j} + \tau_{i,j}^{(2)} \eta_{i,j}^2-\tau_{z\,min}}{\Delta z}\right)s_z\right)^2 }\\
& +  {{{v_0}}^2}\,{\left( \cos\theta \left(\frac{\tau_{i,j}^{(0)} + \tau_{i,j}^{(1)} \eta_{i,j} + \tau_{i,j}^{(2)} \eta_{i,j}^2-\tau_{z\,min}}{\Delta z}\right)s_z-\sin\theta \left(\frac{\tau_{i,j}^{(0)} + \tau_{i,j}^{(1)} \eta_{i,j} + \tau_{i,j}^{(2)} \eta_{i,j}^2-\tau_{x\,min}}{\Delta x}\right)s_x \right)^2}\\
& \times \left( 1 - 2 \eta_{i,j} {v_{nmo}^2} \,{ \left( \cos\theta \left(\frac{\tau_{i,j}^{(0)} + \tau_{i,j}^{(1)} \eta_{i,j} + \tau_{i,j}^{(2)} \eta_{i,j}^2-\tau_{x\,min}}{\Delta x}\right)s_x +\sin\theta \left(\frac{\tau_{i,j}^{(0)} + \tau_{i,j}^{(1)} \eta_{i,j} + \tau_{i,j}^{(2)} \eta_{i,j}^2-\tau_{z\,min}}{\Delta z}\right)s_z \right)^2} \right)=1.
\end{aligned}
$}
\label{eq:discretettiexpanded}
\end{equation}
Then we expand Equation~(\ref{eq:discretettiexpanded}) and collect terms to form polynomial in $\eta_{i,j}$. Ignoring terms with powers of $\eta_{i,j}$ higher than two, we get an equation of the form:
\begin{equation}
f_0\left(\tau_{i,j}^{(0)}\right) + f_1\left(\tau_{i,j}^{(0)},\tau_{i,j}^{(1)}\right)\,\eta_{i,j} + f_2\left(\tau_{i,j}^{(0)},\tau_{i,j}^{(1)},\tau_{i,j}^{(2)}\right)\,\eta_{i,j}^2 = 1.
\label{eq:polynomial}
\end{equation}
where $f_0, f_1,$ and $f_2$ denote coefficients of different powers of $\eta_{i,j}$ and are functions of traveltime coefficients defined inside the parentheses. For the sake of continuity, the complete expressions for these coefficients are given in \ref{app:A}.

Next, we compare the coefficients of different powers of $\eta_{i,j}$ from the left hand side of Equation~(\ref{eq:polynomial}) to those on the right hand side, in succession, starting with the zeroth power. First we solve $f_0\left(\tau_{i,j}^{(0)}\right)=1$ to obtain the expression for the zeroth order traveltime coefficient, which is given as:
\begin{equation}
\tau_{i,j}^{(0)} = \frac{\left(b_0\tau_{x\,min}+c_0\tau_{z\,min}\right)\cos^2\theta + a_0\left(\tau_{x\,min}+\tau_{z\,min}\right) + \left(d_0\tau_{z\,min}+e_0\tau_{x\,min}\right)\sin^2\theta + \sqrt{D_0}}{\left(b_0+c_0\right)\cos^2\theta + 2a_0 + \left(d_0+e_0\right)\sin^2\theta},
\label{eq:tau0}
\end{equation}
where the following definitions have been used for simplification:
\begin{equation}
\begin{aligned}
a_0 & = \left(v_{nmo}^2-v_0^2\right)\Delta x \, \Delta z \sin\theta s_x s_z \cos\theta,\\
b_0 & = \left(v_{nmo} \Delta z\right)^2 , \\
c_0 & = \left(v_0 \Delta x\right)^2 , \\
d_0 & = \left(v_{nmo} \Delta x\right)^2 , \\
e_0 & = \left(v_0 \Delta z\right)^2, \\
D_0 & = \Delta x^2 \, \Delta z^2\left(-v_0^2 v_{nmo}^2 \left(\tau_{x\,min}-\tau_{z\,min}\right)^2\cos^4\theta + 2a_0 \right.\\
&+\left(b_0+c_0-2v_{nmo}^2v_0^2 \left(\tau_{x\,min}-\tau_{z\,min}\right)^2 \sin^2\theta \right)\cos^2\theta\\
& \left. + \left(d_0 + e_0 -v_{nmo}^2v_0^2\left(\tau_{x\,min}-\tau_{z\,min}\right)^2 \sin^2\theta\right)\sin^2\theta \right).
\end{aligned}
\end{equation}
Note that $\tau_{i,j}^{(0)}$ is the solution of Equation~(\ref{eq:discretetti}) when $\eta_{i,j}=0$, which would be the discretized form of the TEA eikonal equation. Hence, like~\cite{stovas_new_2012} the zeroth order coefficient is the solution to the TEA eikonal equation.

Next, solving $f_1\left(\tau_{i,j}^{(0)},\tau_{i,j}^{(1)}\right)=0$ results in the expression for $\tau_{i,j}^{(1)}$ in terms of the already computed traveltime $\tau_{i,j}^{(0)}$. This is given as:
\begin{equation}
\begin{aligned}
\tau_{i,j}^{(1)} = & -\frac{1}{D_1}\biggl(v_{nmo}^2\left(b_1 s_x\cos\theta + a_1 s_z\sin\theta\right)^2\left(b_1 v_0 s_x\sin\theta + \Delta x \Delta z -a_1 v_0 s_z \cos\theta\right)\biggr. \\
& \times \biggl. \left(a_1 v_0 s_z\cos\theta + \Delta x \Delta z -b_1 v_0 s_x \sin\theta\right)\biggr),
\label{eq:tau1}
\end{aligned}
\end{equation}
where
\begin{equation}
\begin{aligned}
a_1 & = \Delta x \, \left(\tau^{(0)}_{i,j} - \tau_{z\,min}\right), \\
b_1 & = \Delta z \, \left(\tau^{(0)}_{i,j} - \tau_{x\,min}\right), \\
c_1 & = \left(v_{nmo}^2-v_0^2\right)\Delta x \, \Delta z \, \left(2\tau^{(0)}_{i,j} - \tau_{x\,min} - \tau_{z\,min}\right),\\
D_1 & = \Delta x^2 \, \Delta z^2 \left( \left( v_{nmo}^2 \Delta z \, b_1 + v_0^2 \, \Delta x \, a_1 \right) \cos^2\theta \right. \\
 & \left. + c_1 \cos\theta \sin\theta + \left(v_{nmo}^2 \Delta x \, a_1 + v_0^2 \, \Delta z \, b_1\right) \sin^2\theta \right).
\end{aligned}
\end{equation}

\noindent Finally, solving the equation $f_2\left(\tau_{i,j}^{(0)},\tau_{i,j}^{(1)},\tau_{i,j}^{(2)}\right)=0$ yields the expression for $\tau_{i,j}^{(2)}$ in terms of the known traveltime coefficients $\tau_{i,j}^{(0)}$ and $\tau_{i,j}^{(1)}$ as follows:
\begin{equation}
\resizebox{0.8\hsize}{!}{$
\begin{aligned}
\tau_{i,j}^{(2)}  & =  -\frac{1}{D_2} \left(\tau_{i,j}^{(1)}\left(-4 \Delta x^2 \Delta z^2 a_2 b_2 c_2 v_{nmo}^2 v_0^2 \cos^4\theta - 4\Delta x \Delta z v_{nmo}^2 v_0^2 s_x s_z \right. \right.\\
& \times \left(-\Delta z^2 a_2^2 d_2 + \Delta x^2 b_2^2 e_2\right) \cos^3 \theta \sin\theta + \Delta x^2 \Delta z^2 \sin^2 \theta \left( \Delta z^2 \tau_{i,j}^{(1)} v_0^2 + 4\Delta x^2 \tau_{i,j}^{(0)} v_{nmo}^2 \right.\\
& \left. + \Delta x^2 \tau_{i,j}^{(1)} v_{nmo}^2 - 4\Delta x^2 \tau_{z\,min} v_{nmo}^2 - 4a_2 b_2 c_2 v_{nmo}^2 v_0^2 \sin^2 \theta \right) + 2\Delta x \Delta z s_x s_z \cos\theta \sin\theta \\
& \times \left(\Delta x^2 \Delta z^2 \left(4\tau_{i,j}^{(0)} v_{nmo}^2 -2\left(\tau_{x\,min} + \tau_{z\,min}\right) + \tau_{i,j}^{(1)}\left(v_{nmo}^2 - v_0^2\right)\right) - 2 v_{nmo}^2 v_0^2 \right.\\
& \times \left. \left(\Delta z^2 a_2^2 d_2 - 
\Delta x^2 b_2 e_2\right) \sin^2\theta \right) + \cos^2\theta \left(\Delta x^2 \Delta z^2 \left(\Delta z^2 \left(4\tau_{i,j}^{(0)} + \tau_{i,j}^{(1)}-4 \tau_{x\,min}\right) v_{nmo}^2 \right. \right. \\
 & + \left. \left. \left. \left. \Delta x^2 \tau_{i,j}^{(1)} v_0^2 \right) - 8v_{nmo}^2 v_0^2 \left(\Delta z^4 a_2^3 -2 \Delta x^2 \Delta z^2 a_2 b_2 c_2 + \Delta x^4 b_2^3\right) \sin^2\theta \right)\right) \right),
\end{aligned}
$}
\label{eq:tau2}
\end{equation}
where
\begin{equation}
\begin{aligned}
a_2 & = \tau_{i,j}^{(0)}-\tau_{x\,min},\\
b_2 & = \tau_{i,j}^{(0)}-\tau_{z\,min}, \\
c_2 & = 2\tau_{i,j}^{(0)}-\tau_{x\,min} -\tau_{z\,min}, 
\\ d_2 & = 4\tau_{i,j}^{(0)}-\tau_{x\,min} -3\tau_{z\,min}, \\ 
e_2 & = 4\tau_{i,j}^{(0)} - 3\tau_{x\,min} -\tau_{z\,min}\\
D_2 & = 2 \Delta x^2 \, \Delta z^2 \left(\left( \Delta z^2 \, a_2 v_{nmo}^2 + \Delta x^2 \, b_2 v_0^2 \right) s_x s_z \cos^2\theta + \Delta x \, \Delta z c_2 \left(v_{nmo}^2-v_0^2\right)\cos\theta \sin\theta \right. \\
& + \left. \left(\Delta x^2 \, b_2 v_{nmo}^2 + \Delta z^2 \, a_2 v_0^2\right)\sin^2\theta\right)
\end{aligned}
\end{equation}

Following this way, we can derive expressions for traveltime coefficients corresponding to terms higher than the second order as well in the perturbation expansion.

For the case of TI media with vertical symmetry axis (VTI), i.e. $\theta=0^{\circ}$, the traveltime coefficients given by equations~(\ref{eq:tau0}),~
(\ref{eq:tau1}), and~(\ref{eq:tau2}) reduce to:
\begin{equation}
\begin{aligned}
\tau_{i,j}^{(0)} & = \frac{1}{\Delta x^2 v_0^2+\Delta z^2 v_{nmo}^2}\biggl(\Delta x^2 \tau_{z\,min} v_0^2+\Delta z^2 \tau_{x\,min} v_{nmo}^2 \biggr.\\
& + \biggl. \sqrt{\Delta x^2 \Delta z^2 \left(v_0^2 \left(\Delta x^2-v_{nmo}^2 (\tau_{x\,min}-\tau_{z\,min}^2\right)+\Delta z^2
   v_{nmo}^2\right)}\biggr), \hspace{2.2cm}
\label{eq:tau0vti}
\end{aligned}
\end{equation}
\begin{equation}
\tau_{i,j}^{(1)} = \frac{v_{nmo}^2 \left(\tau_{i,j}^{(0)}-\tau_{x\,min}\right)^2 \left(v_0^2 \left(\tau_{i,j}^{(0)}-\tau_{z\,min}\right)^2-\Delta z^2\right)}{\Delta x^2 v_0^2
\left(\tau_{i,j}^{(0)}-\tau_{z\,min}\right)+\Delta z^2 \nu_{nmo}^2 \left(\tau_{i,j}^{(0)}-\tau_{x\,min}\right)}, \hspace{3.1cm}
\label{eq:tau1vti}
\end{equation}
\begin{equation}
\begin{aligned}
\tau_{i,j}^{(2)}  & = \frac{-1}{2 \left(\Delta x^2 v_0^2
\left(\tau_{i,j}^{(0)}-\tau_{i,j-1}\right)+\Delta z^2 v_{nmo}^2 \left(\tau_{i,j}^{(0)}-\tau_{x\,min}\right)\right)}\biggl(\tau_{i,j}^{(1)} \left(v_0^2 \left(\Delta x^2 \tau_{i,j}^{(1)} \right. \right. \biggr.\\
&- \left. 4 v_{nmo}^2 \left(\tau_{i,j}^{(0)}-\tau_{x\,min}\right) \left(\tau_{i,j}^{(0)}-\tau_{z\,min}\right) \left(2
\tau_{i,j}^{(0)}-\tau_{x\,min}-\tau_{z\,min}\right)\right)\\
&+ \biggl. \left. \Delta z^2 v_{nmo}^2 \left(4 \tau_{i,j}^{(0)}+\tau_{i,j}^{(1)}-4 \tau_{x\,min}\right)\right)\biggr),  \hspace{5cm}
\label{eq:tau2vti}
\end{aligned}
\end{equation}
respectively.

\subsection{Accelerating the convergence rate}
\noindent Shanks transform~\cite{bender_advanced_1999} can be used to increase the rate of convergence and obtain an even more accurate approximation, once the first three terms of the expansion~(\ref{eq:trial}) have been evaluated. The first sequence of Shanks transform, employing traveltime coefficients $\tau_{i,j}^{(0)}, \tau_{i,j}^{(1)},$ and $\tau_{i,j}^{(2)}$ is given as:
\begin{equation}
\tau_{i,j} \approx \tau_{i,j}^{(0)}+\frac{\eta_{i,j} \left(\tau_{i,j}^{(1)}\right)^2}{\tau_{i,j}^{(1)}-\eta_{i,j} \tau_{i,j}^{(2)}}.
\label{eq:shanks}
\end{equation}

It is also possible to create a sequence of Shanks transform estimates and apply the Shanks transform on them. For example, if the first five traveltime coefficients $\tau_{i,j}^{(0)}, \tau_{i,j}^{(1)}, \tau_{i,j}^{(2)}, \tau_{i,j}^{(3)},$ and $\tau_{i,j}^{(4)}$ have been computed, then one can define a sequence of Shanks transform $S_1, S_2, S_3$, where:
\begin{equation}
\begin{aligned}
S_1 & \approx \tau_{i,j}^{(0)}+\frac{\eta_{i,j} \left(\tau_{i,j}^{(1)}\right)^2}{\tau_{i,j}^{(1)}-\eta_{i,j} \tau_{i,j}^{(2)}}, \\
S_2 & \approx \tau_{i,j}^{(1)}+\frac{\eta_{i,j} \left(\tau_{i,j}^{(2)}\right)^2}{\tau_{i,j}^{(2)}-\eta_{i,j} \tau_{i,j}^{(3)}}, \\
S_3 & \approx \tau_{i,j}^{(2)}+\frac{\eta_{i,j} \left(\tau_{i,j}^{(3)}\right)^2}{\tau_{i,j}^{(3)}-\eta_{i,j} \tau_{i,j}^{(4)}}. \\
\label{eq:shanks1}
\end{aligned}
\end{equation}

Then the Shanks transform formula can be used on these Shanks estimates to obtain an even accurate estimate of traveltime:
\begin{equation}
\tau_{i,j} \approx \frac{S_1 S_3 -S_2^2}{S_1 -2S_2 + S_3}.
\label{eq:shanks2}
\end{equation}

In seismic exploration, the values of $\eta$ observed in practice are close to zero, therefore for numerical tests we will restrict our computations to the first sequence of Shanks transform that requires evaluation of the first three terms in the perturbation expansion. However, for other domains of computational physics that require a solution to the anisotropic eikonal equation, Equation~(\ref{eq:shanks2}) can be employed effectively, in case the value of the perturbed parameter is significantly far from the point of expansion.

\section{Algorithm}
\label{sec:3}

The numerical algorithm is based on the fast sweeping scheme. It is an efficient iterative method that uses Gauss-Seidel iterations with different orderings to solve a wide range of Hamilton-Jacobi equations~\cite{boue_markov_1999}. With an appropriate upwind difference scheme that ensures causality of the underlying PDE, the algorithm converges in a finite number of iterations independent of the mesh size (proved in~\cite{zhao_fast_2005}). By using many sweeping iterations with alternate orderings, it is ensured that for every grid point at least one sweeping direction matches the characteristic direction of the underlying PDE. The value at every grid node is non-decreasing during the sweeps due to the update rule. The correct value for a grid point is reached when it obtains the minimal value it can attain and this value is not changed in the later iterations.

Let $\bar{\tau}$ be the final traveltime solution that we seek and $\tau$ be the intermediary solution. The algorithm begins by assigning an extremely large positive traveltime value to all grid points. Then, the whole domain is swept repeatedly with four alternating orderings. Within each sweep, the main computation and update algorithm is executed (see Algorithm~\ref{alg:main}).\\

\begin{algorithm}\caption{Main Loop}
\label{alg:main}
\begin{algorithmic}

\State $\tau, \bar{\tau} \Leftarrow \infty$  \Comment{Set an infinitely large value for traveltimes}
\Repeat  \Comment{Gauss-Seidel iterations with alternating sweeping order}
\For{$i = 1 \to I, j=1 \to J$}
\State \textit{run the inner loop}
\EndFor

\For{$i = 1 \to I, j=J \to 1$}
\State \textit{run the inner loop}
\EndFor

\For{$i = I \to 1, j=1 \to J$}
\State \textit{run the inner loop}
\EndFor

\For{$i = I \to 1, j=J \to 1$}
\State \textit{run the inner loop}
\EndFor

\Until{convergence}

\end{algorithmic}
\end{algorithm}

For every grid point $(i,j)$, the computations begin by evaluating the minimum valued neighbors according to Equation~(\ref{eq:neighbor}). Depending on the neighbor chosen, sign variables $s_x, s_z$ are evaluated as per Equation~(\ref{eq:sign}). This ensures that an upwind discretization is used. Then, we compute the traveltime coefficients $\tau_{i,j}^{(0)}, \tau_{i,j}^{(1)},$ and $\tau_{i,j}^{(2)}$ using Equations~(\ref{eq:tau0}),~(\ref{eq:tau1}), and~(\ref{eq:tau2}), respectively. Once we have the first three coefficients of the perturbation expansion, we can use the first sequence of Shanks transform, given by Eq.~(\ref{eq:shanks}), to obtain a better approximation to the traveltime expansion. This traveltime solution, $\tau_{i,j}$, will be accepted if it satisfies the causality criterion (see \ref{app:B}) and is smaller than the value already stored in $\bar{\tau}_{i,j}$. In that case, $\bar{\tau}_{i,j}$ will be update and the algorithm will return to the main loop, otherwise it will return without updating $\bar{\tau}_{i,j}$ (see Algorithm~\ref{alg:inner}).

\begin{algorithm}\caption{Inner loop}
\label{alg:inner}
\begin{algorithmic}

\State \textit{Compute:}

\begin{itemize}
\State  $\tau_{x\,min}, \tau_{z\,min}$ using Equation~(\ref{eq:neighbor}) \Comment{Minimum valued neighbors}
\State $s_x, s_z$ using Equation~(\ref{eq:sign}) \Comment{Sign variables}
\State $\tau_{i,j}^{(0)}$ using Equation~(\ref{eq:tau0}) \Comment{Zeroth order coefficient}
\State $\tau_{i,j}^{(1)}$ using Equation~(\ref{eq:tau1}) \Comment{First order coefficient}
\State $\tau_{i,j}^{(2)}$ using Equation~(\ref{eq:tau2}) \Comment{Second order coefficient}
\State $\tau_{i,j}$ using Equation~(\ref{eq:shanks}) \Comment{Shanks transform}
\end{itemize}

\If {$\tau_{i,j}$ is causal and $\tau_{i,j}<\bar{\tau}_{i,j}$}  
\State $\bar{\tau}_{i,j} \Leftarrow \tau_{i,j}$ \Comment{Traveltime update}
\EndIf
\State \textit{return to the main loop}

\end{algorithmic}
\end{algorithm}

\section{Numerical Tests}
\label{sec:4}

In this section we test our numerical algorithm on a variety of examples and demonstrate its accuracy and efficiency properties. We will begin with a homogeneous TTI model test. Even though the practical implications for such a model are limited, however, it is extremely useful for studying the error and convergence properties of the algorithm. Then we will test our algorithm on benchmark synthetic models: the VTI Marmousi model~\cite{alkhalifah_anisotropic_1997}, and a section of the BP TTI model~\cite{shah_2007}. These tests are likely indicators of how the algorithm will perform in practical scenario.

\begin{figure}[ht!]
\begin{center}
\subfigure[]{%
\label{hom1}
\includegraphics[width=0.4\textwidth]{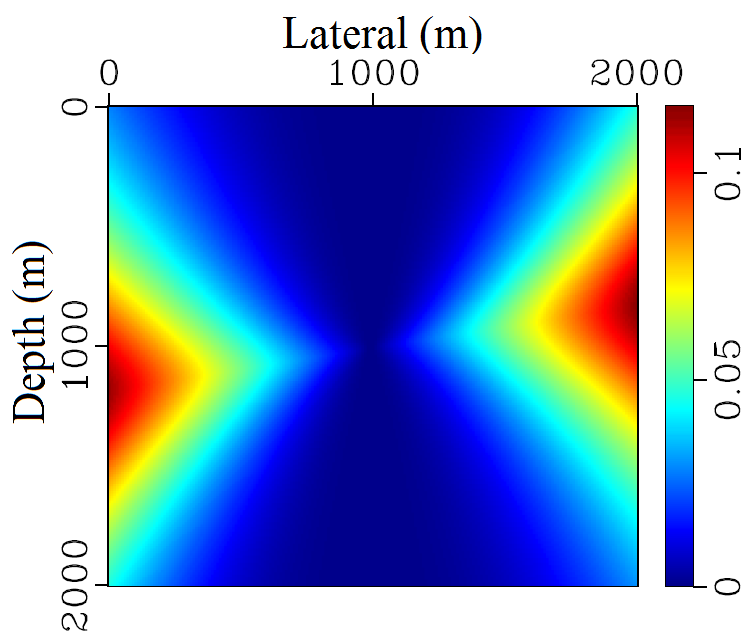}
}
\subfigure[]{%
\label{hom2}
\includegraphics[width=0.4\textwidth]{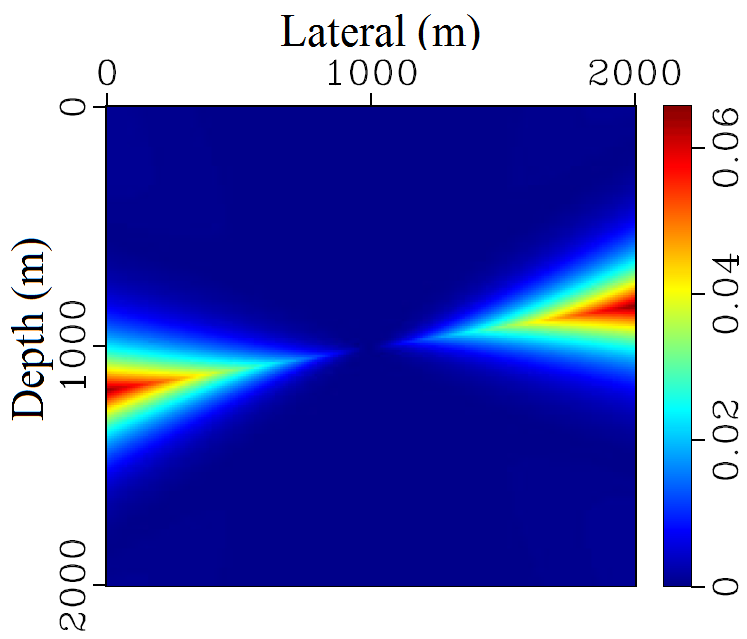}
}\\
\subfigure[]{%
\label{hom3}
\includegraphics[width=0.4\textwidth]{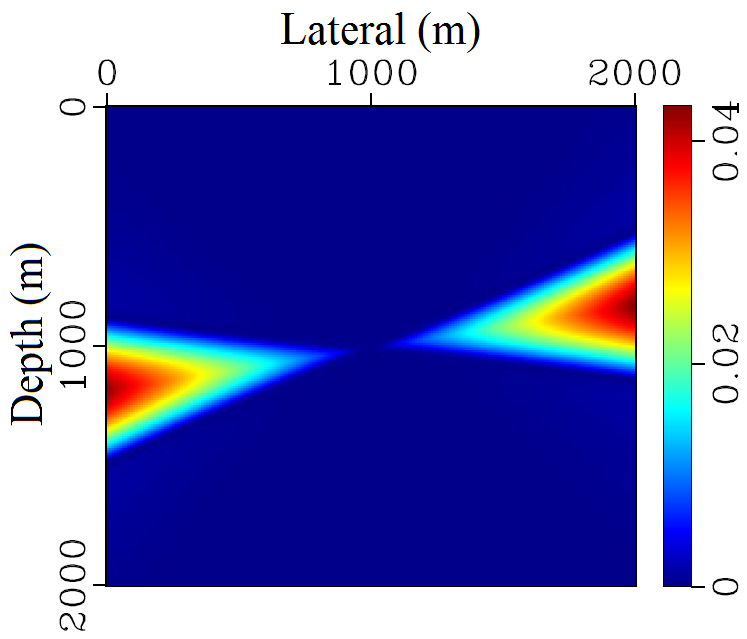}
}
\subfigure[]{%
\label{hom4}
\includegraphics[width=0.4\textwidth]{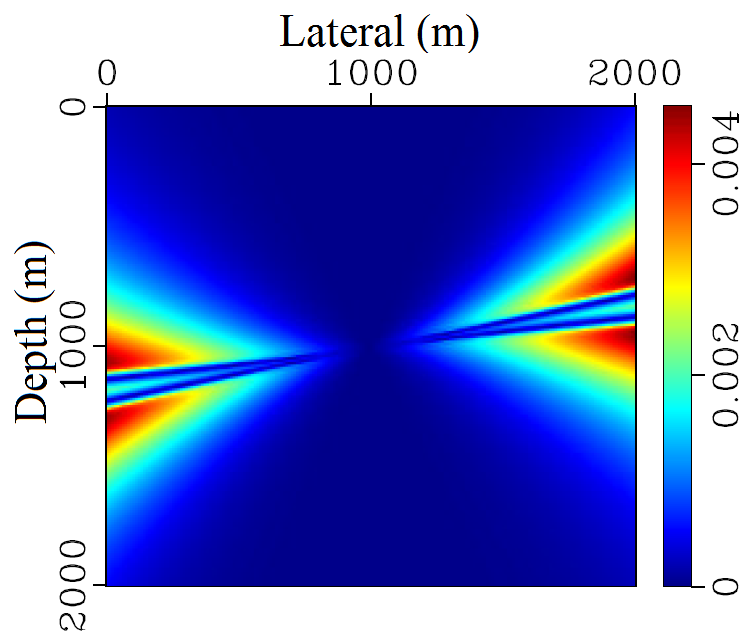}
}
\end{center}
\caption{
Absolute traveltime errors in seconds for a homogeneous TTI model with $v_0=2$ km/s, $v_{nmo} = 2.2$ km/s, $\eta=0.4$, and $\theta=10^{\circ}$ considering up to zeroth order (a), first order (b), second order (c) terms of the expansion and the first sequence of Shanks transform (d). The grid spacing used is $\Delta x = \Delta z = 10$ m and the source is located at (1000 m, 1000 m).
}
\label{hom}
\end{figure}

{\bf Example 1. Homogeneous TTI model :} First we consider a $2$ km $\times$ $2$ km homogeneous TTI model with $v_0 = 2$ km/s, $v_{nmo} = 2.2$ km/s, $\theta = 10^{\circ}$, and $\eta = 0.4$. The value of $\eta$ is deliberately chosen to be significantly higher than those encountered in practice. Therefore, we can consider the error property of the algorithm for practical cases to be bounded by this test case. The grid spacing used in both directions is $10$~m. We place a source at the centre of the model $(1000$ m, $1000$ m) and compute traveltimes using the proposed algorithm for the whole model.

For comparison, we also compute the solution to the discretized eikonal equation~(\ref{eq:discretetti}) by solving a quartic polynomial using Bairstow's algorithm~\cite{ibanez_effective_2013}. In this case, the second largest real root corresponds to the correct outgoing P-wave solution~\cite{qian_paraxial_2001}. This will be used as the reference solution in all the experiments and will be referred to as `the direct solver'.

Figure~\ref{hom} shows absolute traveltime errors for our proposed algorithms involving different terms in the perturbation expansion. In Figure~\ref{hom1}, we consider only the zeroth order term $\left(\tau_{i,j}^{(0)}\right)$ of the expansion, which is the solution to TEA eikonal equation. Hence, Figure~\ref{hom1} shows the traveltime errors caused by ignoring the $\eta$ parameter. The maximum effect of the $\eta$ parameter is on the wave propagating in the direction orthogonal to the symmetry axis. Hence, we observe that the maximum errors are generated at $10^{\circ}$ from the horizontal axis. The peak error in Fig.~\ref{hom1} is 116.2 ms.

Then, we include the first order term $\left(\tau_{i,j}^{(1)}\right)$ as well in the expansion and plot absolute traveltime errors in Figure~\ref{hom2}. As expected, the errors reduce compared to Figure~\ref{hom1}. However, still the maximum error is quite significant $(65.7$ ms). Next, we incorporate the second order term $\left(\tau_{i,j}^{(2)}\right)$ into the perturbation expansion and plot absolute traveltime errors in Figure~\ref{hom3}. The maximum error reduces to 43.2 ms when we use the first three terms of the traveltime expansion.

At this point, we can observe a clear trend of reduction in errors as we continue to add terms into the traveltime expansion [Equation~(\ref{eq:trial})]. It is evident that we might need to compute many terms of the expansion before obtaining sufficiently accurate traveltimes. However, since we have already computed the first three terms of the expansion, we can use the first sequence of Shanks transform to accelerate the convergence rate. Figure~\ref{hom4} plots traveltime errors for the Shanks transform based traveltime expression, given by Equation~(\ref{eq:shanks}). Notice that the convergence rate dramatically increases causing a significant reduction of traveltime errors. The advantage of using Shanks transform is that this improvement in accuracy is obtained with merely two addition and four multiplication operations, a significantly lower cost than computing any of the higher order terms. The peak traveltime error reduces to only 4.5 ms in Figure~\ref{hom4}. This low error value is extremely encouraging, considering that an unrealistically large value of $\eta$ was used. One can follow the methodology outlined in Section 2 and obtain expressions for higher order traveltime coefficients and use them to obtain even more accurate traveltimes.

Table~\ref{tbl:efficiency} shows the cost associated with the proposed algorithm as percentage of the cost needed by the direct solver. We note that the zeroth order term or the TEA solution costs only 17.7~\% of the TTI solution using the direct solver. The difference in cost is mainly associated with the fact that the TEA eikonal equation yields a quadratic polynomial in the finite difference scheme, for which there exists a stable closed-form solution. However, for a quartic polynomial obtained for the TTI eikonal equation, the analytical solution is numerically instable~\cite{higham_accuracy_1996} and hence requires a costlier numerical root finding algorithm.

Even though the TEA solution is computationally cheaper, however, the accuracy is significantly low as well. Therefore, we need to include additional terms of the expansion in order to improve the traveltime accuracy. We observe that as we include further terms, the computational cost rises. However, the rate at which the cost increases is significantly lower than the rate of error reduction. This is mainly due to the fact that we obtained analytical expressions for traveltime coefficients, which is much cheaper to compute than numerical root finding algorithms. Finally, we see that at a minor increment in cost, we can use the first sequence of Shanks transform to increase the convergence rate significantly. The computations were performed using C code on a 2.4 GHz Intel Pentium machine with 4GB of RAM.

\begin{table}
\centering
{\tabulinesep=1.9mm
\begin{tabu}{|c|c|}
\hline
{\bf \large Order} & {\bf \large Cost} \\ \hline
Zeroth & 17.7 \% \\ \hline
First & 18.7 \% \\ \hline
Second & 20.4 \% \\ \hline
Shanks $(S_1)$ & 21.1 \% \\ \hline
\end{tabu}}
\caption{Computational cost associated with computing up to different terms in the perturbation expansion. The costs are given as percentage of the cost needed by the direct solver.
\label{tbl:efficiency}}
\end{table}

In the following examples we will test only the traveltime expansion based on the first sequence of Shanks transform $(S_1)$ on complex anisotropic models and evaluate its accuracy compared to the direct solver.


{\bf Example 2. The VTI Marmousi model :} In this example, we test the accuracy of the algorithm on the famous VTI Marmousi model. The geometry of the Marmousi model is based on a profile through the North Quenguela trough in the Cuanza basin~\cite{versteeg_sensitivity_1993}. The anellipticity parameter $\eta$ ranges from zero to $0.274$ based on two hypothetical assumptions~\cite{alkhalifah_anisotropic_1997}: 
\begin{itemize}
\item anisotropy increases as velocity decreases (shales have low velocity overall),
\item the horizontal velocity gradually increases with depth.
\end{itemize}

In this model, the vertical velocity $v_0$ is set equal to the NMO velocity $v_{nmo}$. The grid spacing for the model is 12 m in both directions.

\begin{figure}
\begin{center}
\includegraphics[height=0.3\textwidth]{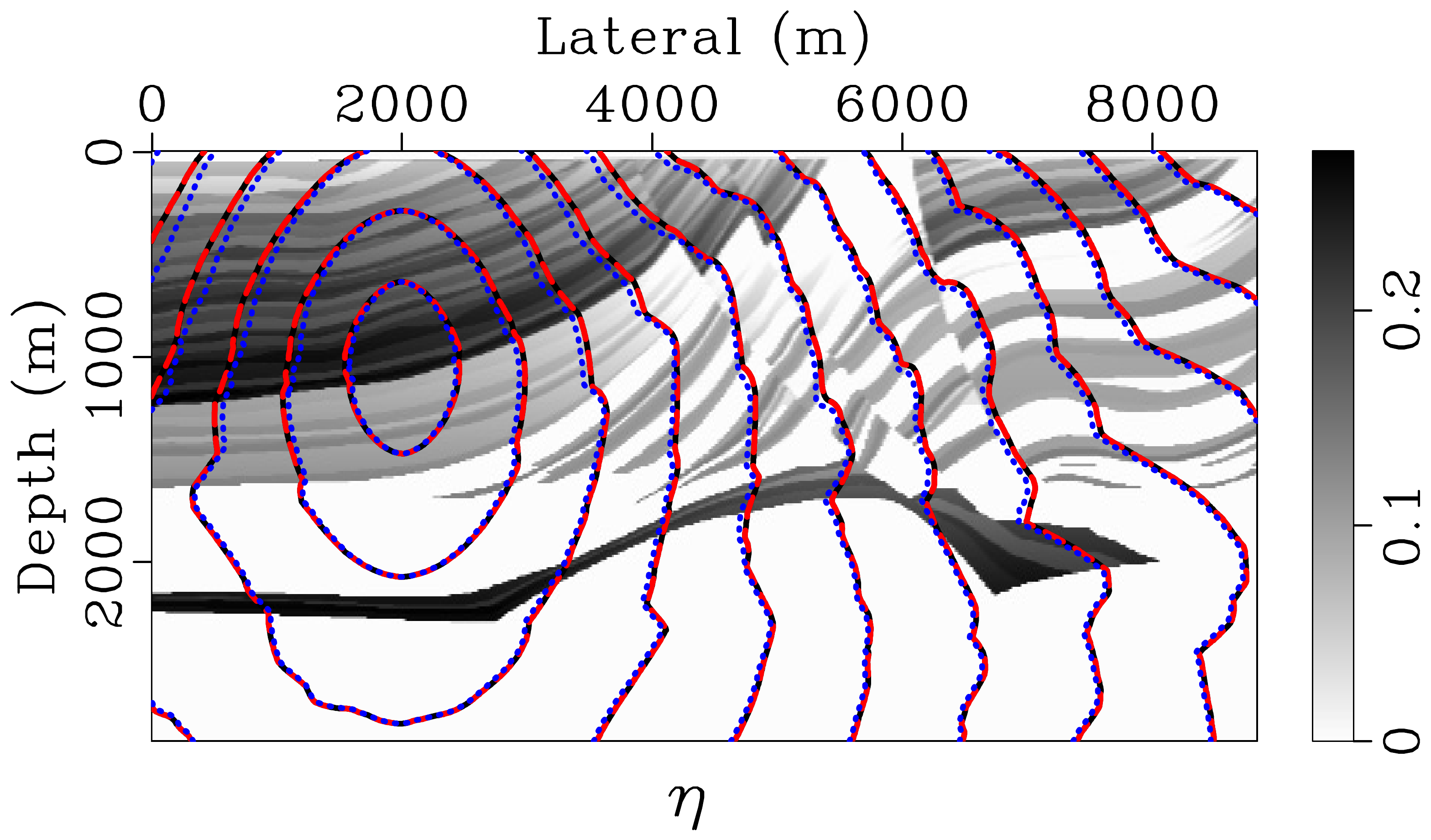}
\caption {Traveltime contours for the VTI Marmousi model mapped on top of the $\eta$ model using the direct solver~(solid black), traveltime expansion based on the first sequence of Shanks transform~(dashed red), and the anelliptic approximation~(dotted blue) by Fomel~\cite{fomel_anelliptic_2004}. A grid spacing of 12 m is used in both directions. The source is located at (2000 m, 1000 m). 
\label{fig:marm1}}
\end{center}
\end{figure}

\noindent In order to stretch the accuracy limits of the algorithm, we place a source in the middle of the large $\eta$ zone, at (2000 m, 1000 m), and compute traveltimes for the whole model. The computed traveltimes are shown in Figure~\ref{fig:marm1} (dashed red contours). For comparison we also plot the reference solution obtained using the direct solver (solid black contours). Notice that a very good match is obtained even in the region with large $\eta$ values. We also plot traveltimes using the anelliptic approximation for VTI media by Fomel~\cite{fomel_anelliptic_2004} (dotted blue contours). The anelliptic approximation costs approximately the same as our proposed algorithm using the first sequence of Shanks transform. However, we could see from Figure~\ref{fig:marm1} that the accuracy of the anelliptic approximation suffers in the region where the effect of $\eta$ is large.

\begin{figure}
\begin{center}
\includegraphics[height=0.3\textwidth]{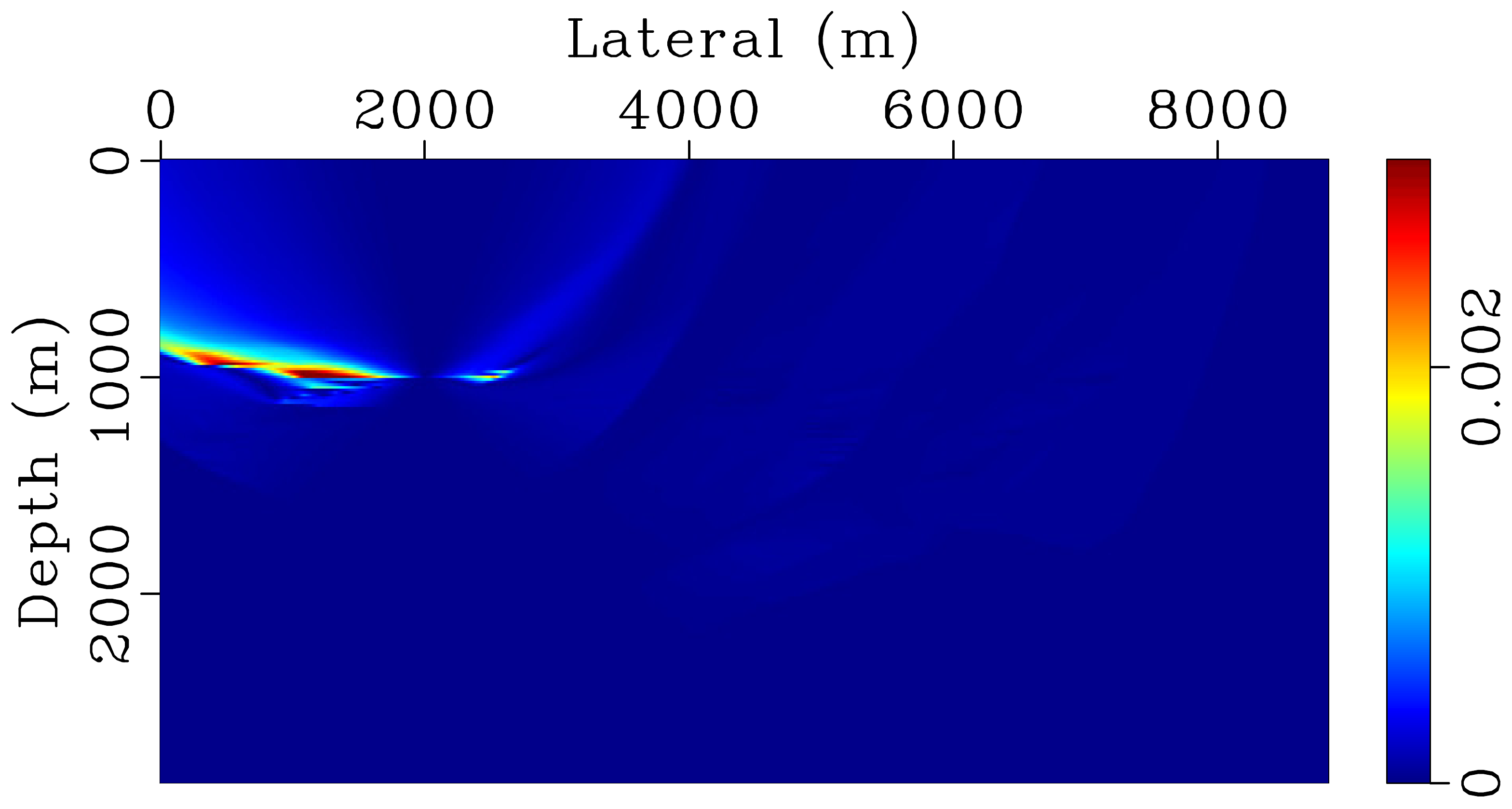}
\caption {Absolute traveltime errors in seconds for the VTI Marmousi using the traveltime expansion based on the first sequence of Shanks transform. A grid spacing of 12 m is used in both directions. The source is located at (2000 m, 1000 m).
\label{fig:marm2}}
\end{center}
\end{figure}

\begin{figure}
\begin{center}
\includegraphics[height=0.3\textwidth]{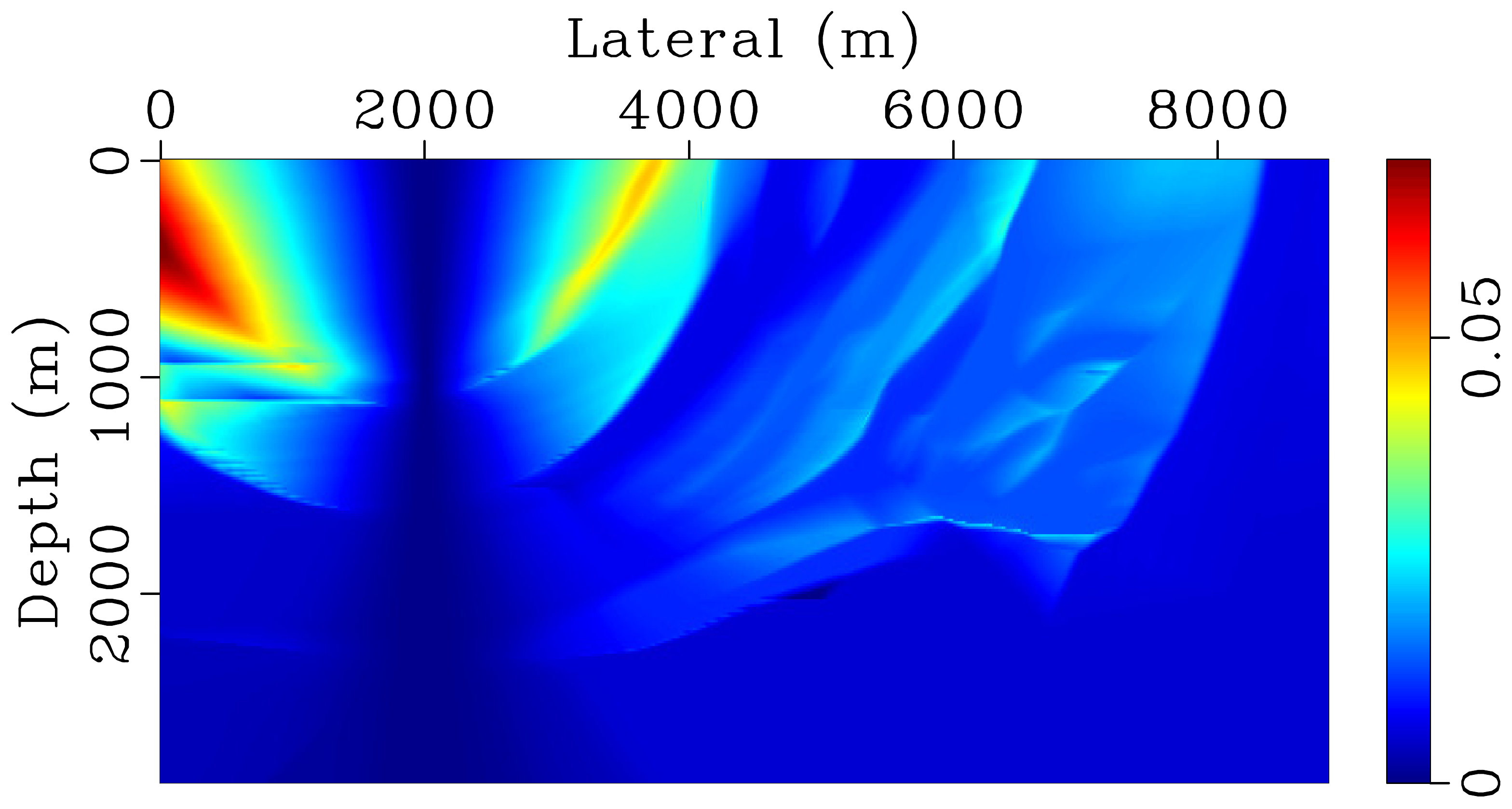}
\caption {Absolute traveltime errors in seconds for the VTI Marmousi using the anelliptic approximation of Fomel~\cite{fomel_anelliptic_2004}.  A grid spacing of 12 m is used in both directions. The source is located at (2000 m, 1000 m).
\label{fig:marm3}}
\end{center}
\end{figure}

We plot the absolute traveltime errors for the proposed algorithm using the first sequence of Shanks transform and the anelliptic approximation in Figures~\ref{fig:marm2}, and~\ref{fig:marm3}, respectively. The solution using the direct solver is used as a reference for both plots. Notice that the maximum error in Figure~\ref{fig:marm2} is merely 3.04~ms compared to a significantly larger value of 69.5~ms in Figure~\ref{fig:marm3}.\\

\begin{figure}[ht!]
\begin{center}
\subfigure[]{%
\includegraphics[width=0.35\textwidth]{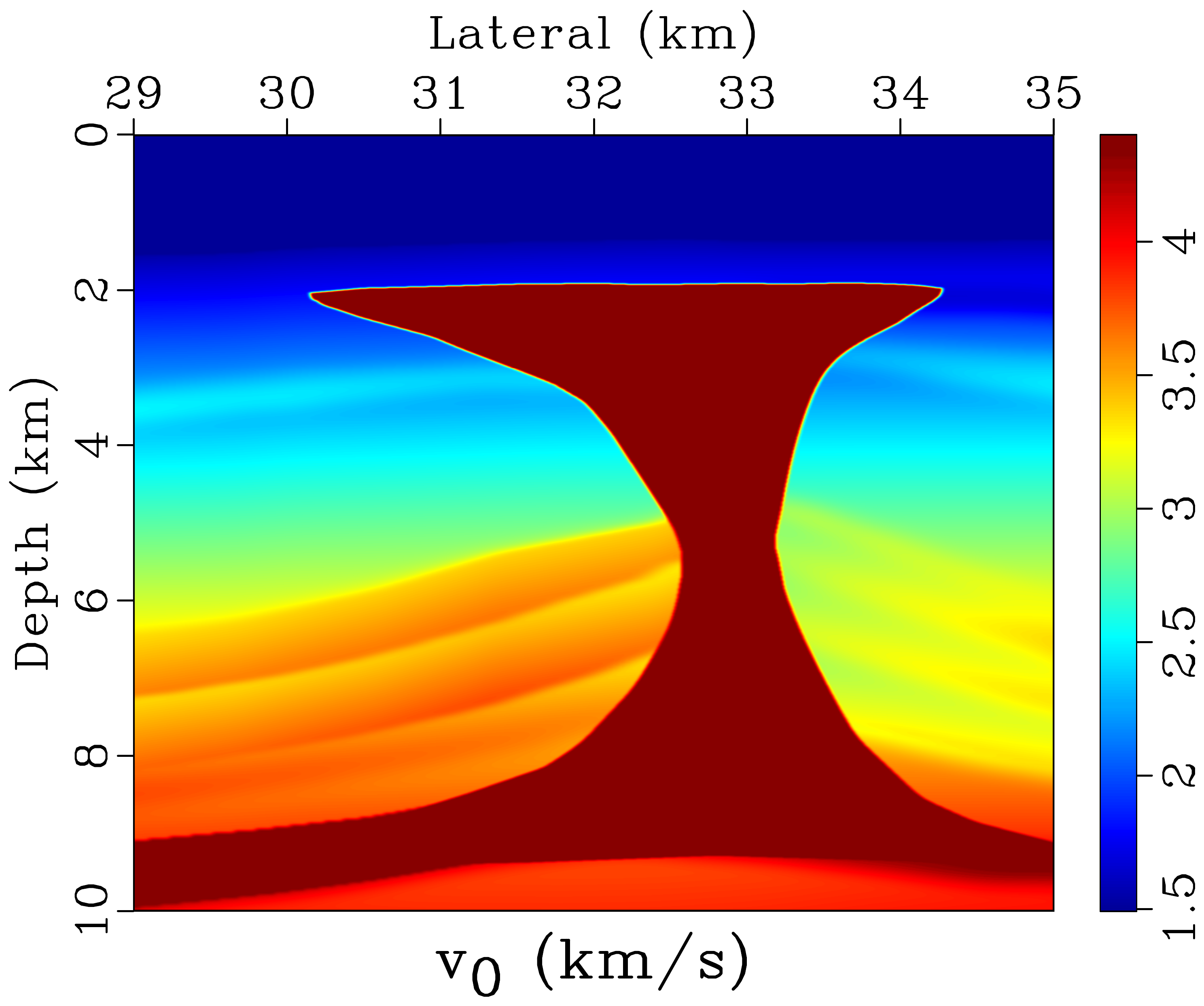}
}
\subfigure[]{%
\includegraphics[width=0.35\textwidth]{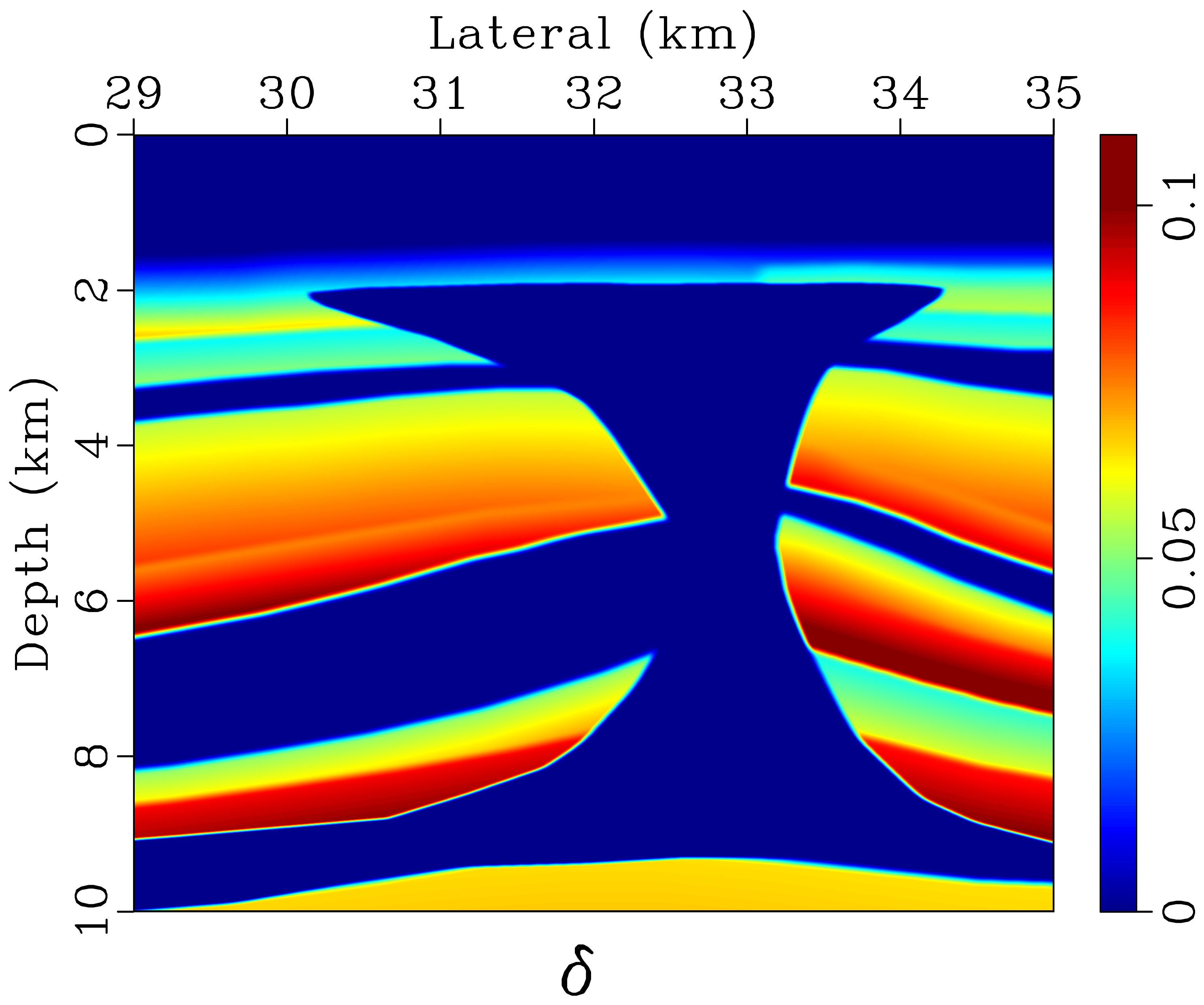}
}\\
\subfigure[]{%
\includegraphics[width=0.35\textwidth]{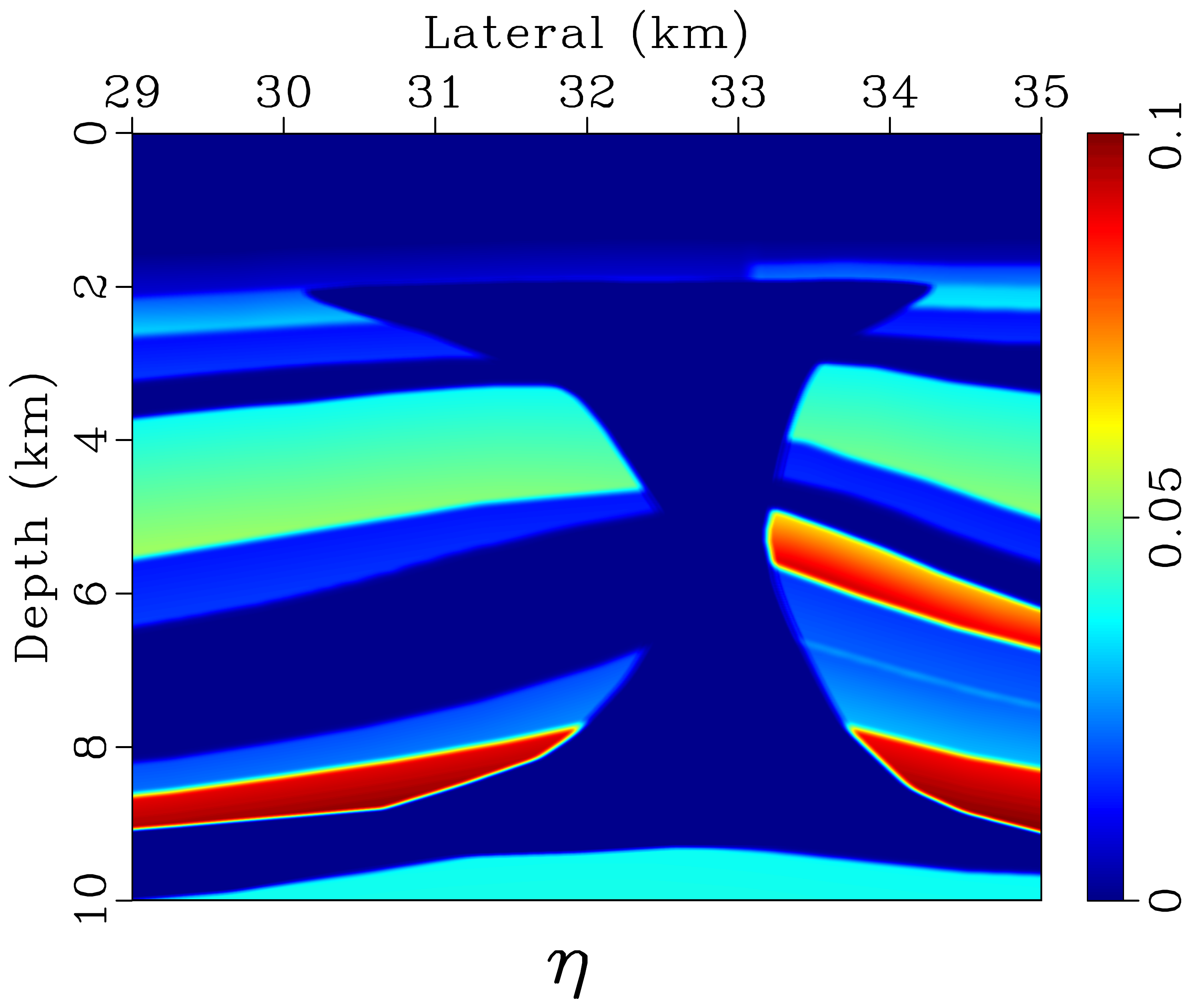}
}
\subfigure[]{%
\includegraphics[width=0.35\textwidth]{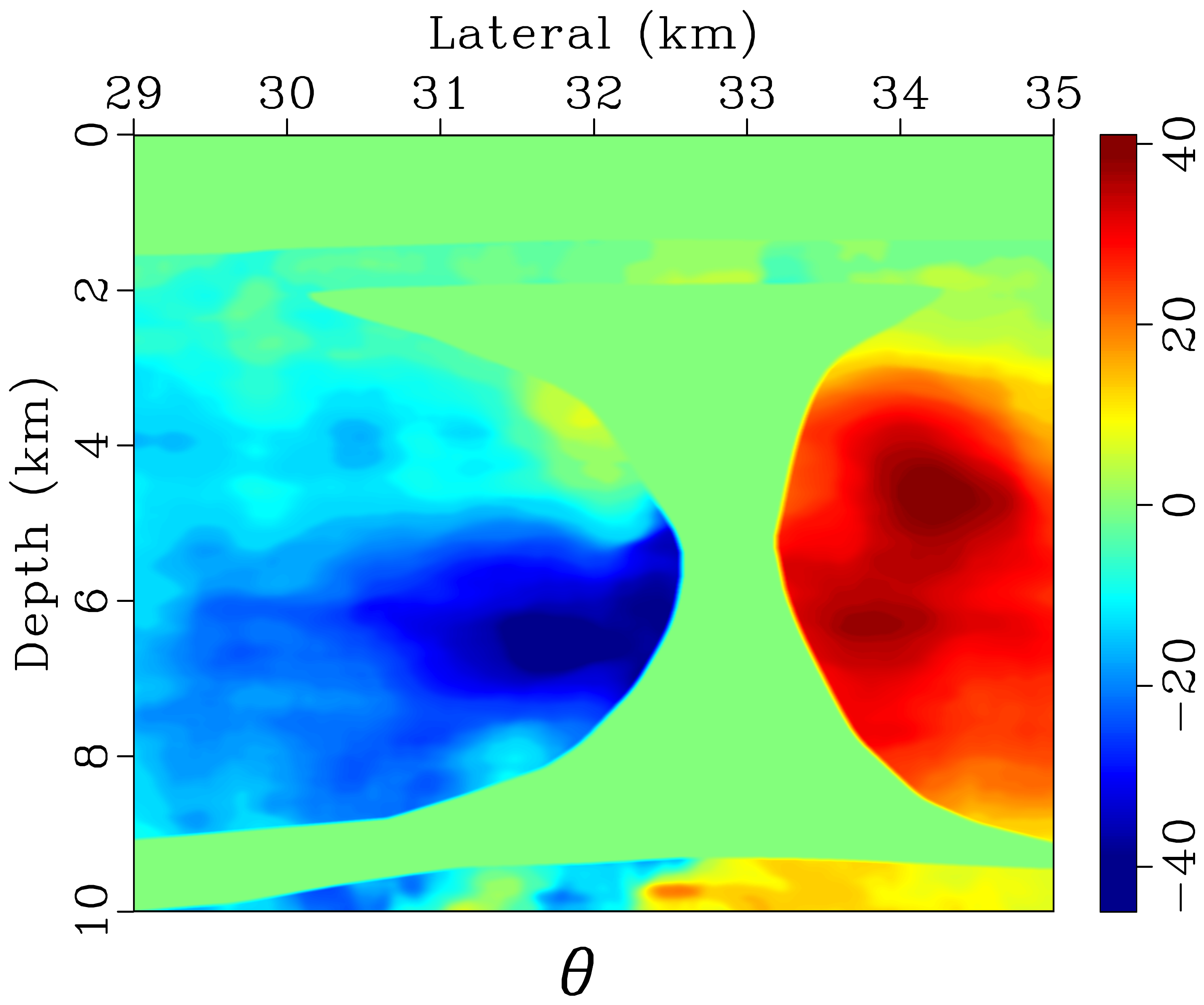}
}
\end{center}
\caption{A portion of the BP TTI model depicting the vertical velocity (a), the NMO velocity (b), the $\eta$ parameter (c), and the tilt (d). 
}%
\label{fig:bpmodel}
\end{figure}

{\bf Example 3. The BP TTI model :} In this section, we test our algorithm on a part of the BP TTI model shown in Figure~\ref{fig:bpmodel}. This section is a key indicator of robustness of anisotropic eikonal solvers due to the sharp variation in tilt values. A grid spacing of 6.25 m is used in both directions.

\begin{figure}
\begin{center}
\includegraphics[height=0.4\textwidth]{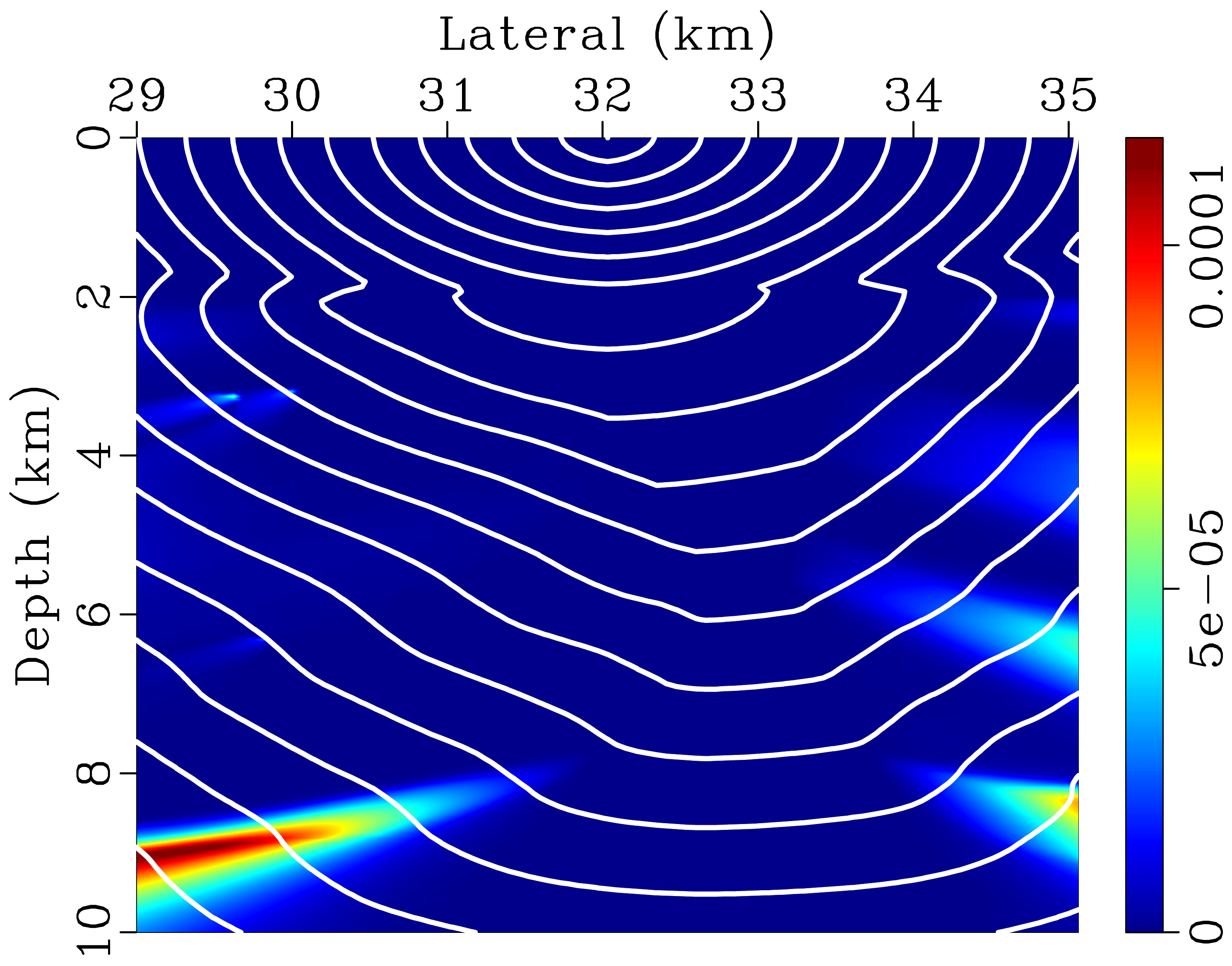}
\caption {Absolute traveltime errors in seconds for a section of the BP TTI model used for the experiment. Also mapped on top are the traveltime contours. The source is located at the center top (32~km,~0~km). A grid spacing of 6.25 m is used in both directions.
\label{fig:bptest}}
\end{center}
\end{figure}

We consider a source at the centre top (32 km, 0 km) and compute solutions using the traveltime expansion based on Shanks transform $(S_1)$ and the direct solver. Figure~\ref{fig:bptest} shows the absolute traveltime errors in seconds obtained for the experiment. Notice that the error is zero almost for the whole domain of experiment, while the maximum error is merely 0.12~ms.

\noindent  

\section{Conclusions}
\label{sec:5}

We developed an efficient algorithm for solving the TTI eikonal equation by utilizing perturbation theory on the first-order discretized form of the TTI eikonal equation. In addition to achieving near perfect accuracy, the proposed algorithm is highly efficient compared to directly solving the TTI eikonal equation. The reduction in cost is obtained due to the derivation of analytic expressions for traveltime coefficients in the perturbation expansion. We also show that Shanks transform can be used to speed up the rate of convergence, requiring the evaluation of only the first few terms in the expansion. Numerical stability is ensured by the use of the fast sweeping method. We demonstrated these assertions through tests on the homogeneous TTI model, the VTI Marmousi model, and the BP TTI model. In addition, we also showed comparison with the anelliptic approximation of Fomel~\cite{fomel_anelliptic_2004}. These tests illustrate the usefulness of the proposed algorithm in traveltime computations for TTI media, where extremely high accuracy can be achieved at a significantly low computational cost.

\section*{Acknowledgments}

We thank KAUST for financial support. We are also grateful David Ketcheson for useful discussions on the direct solver. We also thank BP for releasing the benchmark synthetic model.

\newpage
\appendix

\section{Traveltime functions}
\label{app:A}

In this appendix, we provide complete expressions for the traveltime functions forming the polynomial in $\eta_{i,j}$, shown in Equation~\ref{eq:polynomial}. The coefficient for zeroth power of $\eta_{i,j}$ depends only on the zeroth order traveltime coefficient, $\tau_{i,j}^{(0)}$. The expression for $f_0\left(\tau_{i,j}^{(0)}\right)$ is given as:
\begin{equation}
\begin{aligned}
f_0\left(\tau_{i,j}^{(0)}\right) & =\frac{1}{\Delta x^2 \Delta z^2}\left(\sin^2\theta \left(\Delta x^2 v_{nmo}^2 \left(\tau_{i,j}^{(0)}-\tau_{z\,min}\right)^2+ \Delta z^2 v_0^2 \left(\tau_{i,j}^{(0)}-\tau_{x\,min}\right)^2\right) \right. \\
& +\cos^2\theta \left(\Delta x^2 v_0^2 \left(\tau_{i,j}^{(0)}-\tau_{z\,min}\right)^2 + \Delta z^2 v_{nmo}^2 \left(\tau_{i,j}^{(0)}-\tau_{x\,min}\right)^2\right)-\Delta x^2 \Delta z^2 \\
& \left. +2
\Delta x \Delta z s_x s_z \left(\tau_{i,j}^{(0)}-\tau_{x\,min}\right) \left(\tau_{i,j}^{(0)}-\tau_{z\,min}\right) \sin\theta \cos\theta \left(v_{nmo}^2-v_0^2\right)\right).
\end{aligned}
\end{equation}
The coefficient for the first power of $\eta_{i,j}$ depends on the zeroth and the first order traveltime coefficients $\left(\tau_{i,j}^{(0)}, \tau_{i,j}^{(1)}\right)$ and is given as:
\begin{equation}
\resizebox{0.75\hsize}{!}{$
\begin{aligned}
f_1\left(\tau_{i,j}^{(0)}, \tau_{i,j}^{(1)}\right) & = \frac{2}{\Delta x^4
\Delta z^4} \left(\Delta x \Delta z s_x s_z \sin\theta \cos\theta
   \left(\Delta x^2 \Delta z^2 \left(v_{nmo}^2 \left(2
\tau_{i,j}^{(0)} \tau_{i,j}^{(0)} +2 \tau_{i,j}^{(0)} \right. \right. \right. \right. \\
& \times \left. (\tau_{i,j}^{(1)}-\tau_{x\,min}-\tau_{z\,min})-\tau_{i,j}^{(1)}
(\tau_{x\,min}+\tau_{z\,min})+2 \tau_{x\,min}
   \tau_{z\,min}\right) \\
& + \left. \tau_{i,j}^{(1)} v_0^2 (-2
\tau_{i,j}^{(0)}+\tau_{x\,min}+\tau_{z\,min})\right)-2 v_{nmo}^2 v_0^2 (\tau_{i,j}^{(0)}-\tau_{x\,min})\\
& \left. \times (\tau_{i,j}^{(0)}-\tau_{z\,min}) \sin^2\theta
   \left(\Delta z^2 
(\tau_{i,j}^{(0)}-\tau_{x\,min})^2-\Delta x^2 (\tau_{i,j}^{(0)}-\tau_{z\,min})^2\right)\right)\\
& +\Delta x^2 \Delta z^2
\sin^2\theta \left(\Delta x^2
   v_{nmo}^2 (\tau_{i,j}^{(0)}-\tau_{z\,min})
(\tau_{i,j}^{(0)}+\tau_{i,j}^{(1)}-\tau_{z\,min}) \right. \\
& +\Delta z^2
   \tau_{i,j}^{(1)} v_0^2 
(\tau_{i,j}^{(0)}-\tau_{x\,min})-v_{nmo}^2 v_0^2
   (\tau_{i,j}^{(0)}-\tau_{x\,min})^2
(\tau_{i,j}^{(0)}-\tau_{z\,min})^2 \\
& \left. \times \sin^2\theta \right)-\Delta x^2 \Delta z^2 v_{nmo}^2 v_0^2
(\tau_{i,j}^{(0)}-\tau_{x\,min})^2 (\tau_{i,j}^{(0)}-\tau_{z\,min})^2 \cos^4\theta \\
& +2 \Delta x \Delta z
   v_{nmo}^2 v_0^2 s_x s_z (\tau_{i,j}^{(0)}-\tau_{x\,min})
   (\tau_{i,j}^{(0)}-\tau_{z\,min}) \sin\theta \cos^3\theta
\left(\Delta z^2 \right. \\
& \times \left. (\tau_{i,j}^{(0)}-\tau_{x\,min})^2 -\Delta x^2 (\tau_{i,j}^{(0)}-\tau_{z\,min})^2\right)+\cos^2\theta \left(\Delta x^4 \Delta z^2 \tau_{i,j}^{(1)} v_0^2  \right. \\
& \times (\tau_{i,j}^{(0)}-\tau_{z\,min}) +\Delta x^2 \Delta z^4 v_{nmo}^2 (\tau_{i,j}^{(0)}-\tau_{x\,min})
(\tau_{i,j}^{(0)}+\tau_{i,j}^{(1)}-\tau_{x\,min}) \\
& -v_{nmo}^2 v_0^2 \sin^2\theta \left(\Delta x^4 (\tau_{i,j}^{(0)}-\tau_{z\,min})^4-4
\Delta x^2 \Delta z^2 \right. \\
& \times \left. \left. \left. (\tau_{i,j}^{(0)}-\tau_{x\,min})^2
(\tau_{i,j}^{(0)}-\tau_{z\,min})^2+\Delta z^4
(\tau_{i,j}^{(0)}-\tau_{x\,min})^4\right)\right)\right).
\end{aligned}
$}
\end{equation}
Finally, the coefficient for the second power of $\eta_{i,j}$ depends on the zeroth, the first, and the second order traveltime coefficients $\left(\tau_{i,j}^{(0)}, \tau_{i,j}^{(1)}, \tau_{i,j}^{(2)}\right)$ and is given as:
\begin{equation}
\resizebox{0.75\hsize}{!}{$
\begin{aligned}
f_2\left(\tau_{i,j}^{(0)}, \tau_{i,j}^{(1)}, \tau_{i,j}^{(2)}\right) & = \frac{1}{\Delta x^4 \Delta z^4} \left(\Delta x^2 \Delta z^2 \sin^2\theta \left(\Delta x^2
v_{nmo}^2 \left(4
\tau_{i,j}^{(0)}\tau_{i,j}^{(1)}  +2 \tau_{i,j}^{(0)}\tau_{i,j}^{(2)} \right. \right. \right. \\
& \left. +\tau_{i,j}^{(1)} \tau_{i,j}^{(1)} -2 \tau_{z\,min} (2
\tau_{i,j}^{(1)}+\tau_{i,j}^{(2)})\right) +\Delta z^2 v_0^2 \left(2 \tau_{i,j}^{(0)}
\tau_{i,j}^{(2)} +\tau_{i,j}^{(1)} \tau_{i,j}^{(1)} \right.\\
& \left.  -2 \tau_{i,j}^{(2)} \tau_{x\,min}\right) -4 \tau_{i,j}^{(1)}
v_{nmo}^2 v_0^2 (\tau_{i,j}^{(0)}-\tau_{x\,min})
(\tau_{i,j}^{(0)}-\tau_{z\,min}) \sin^2\theta \\
& \left. \times (2
\tau_{i,j}^{(0)}-\tau_{x\,min}-\tau_{z\,min})\right)+2 \Delta x \Delta z
s_x s_z \sin\theta \cos\theta \left(\Delta x^2 \Delta z^2 \right.\\
& \times \left(v_{nmo}^2  \left(2 \tau_{i,j}^{(0)}(2
\tau_{i,j}^{(1)}+\tau_{i,j}^{(2)})+\tau_{i,j}^{(1)} \tau_{i,j}^{(1)} -2 \tau_{i,j}^{(1)} (\tau_{x\,min}+\tau_{z\,min}) \right. \right.\\
& \left.  -\tau_{i,j}^{(2)}
(\tau_{x\,min}+\tau_{z\,min})\right)+v_0^2 \left(\tau_{i,j}^{(2)} (-2 \tau_{i,j}^{(0)} +\tau_{x\,min}+\tau_{z\,min}) \right.\\
& \left. \left. -\tau_{i,j}^{(1)} \tau_{i,j}^{(1)} \right)\right)+2
\tau_{i,j}^{(1)} v_{nmo}^2 v_0^2 \sin^2\theta (\Delta x^2 (\tau_{i,j}^{(0)}-\tau_{z\,min})^2 \\
& \times \left (4 \tau_{i,j}^{(0)}-3\tau_{x\,min}-\tau_{z\,min})-\Delta z^2 (\tau_{i,j}^{(0)}-\tau_{x\,min})^2 (4
\tau_{i,j}^{(0)}-\tau_{x\,min} \right. \\
& \biggl. \left. -3 \tau_{z\,min})\right)\biggr)-4 \Delta x^2
\Delta z^2 \tau_{i,j}^{(1)} v_{nmo}^2 v_0^2 (\tau_{i,j}^{(0)}-\tau_{x\,min})
(\tau_{i,j}^{(0)}-\tau_{z\,min}) \\
& \times \cos^4\theta (2
\tau_{i,j}^{(0)}-\tau_{x\,min}-\tau_{z\,min}) -4 \Delta x
   \Delta z \tau_{i,j}^{(1)} v_{nmo}^2 v_0^2 s_x
s_z \sin\theta \cos^3\theta\\
&  \left(\Delta x^2 (\tau_{i,j}^{(0)}-\tau_{z\,min})^2 (4 \tau_{i,j}^{(0)}-3
\tau_{x\,min}-\tau_{z\,min})-\Delta z^2 
(\tau_{i,j}^{(0)}-\tau_{x\,min})^2 \right.\\
& \left. \times  (4
\tau_{i,j}^{(0)}-\tau_{x\,min}-3 \tau_{z\,min})\right) +\cos^2\theta \left(\Delta x^4 \Delta z^2 v_0^2 \left(2 \tau_{i,j}^{(0)}
\tau_{i,j}^{(2)} \right. \right.\\
& \left. +\tau_{i,j}^{(1)} \tau_{i,j}^{(1)}-2 \tau_{i,j}^{(2)} \tau_{z\,min}\right) +\Delta x^2 \Delta z^4 v_{nmo}^2 \left(4 \tau_{i,j}^{(0)} \tau_{i,j}^{(1)}+2 \tau_{i,j}^{(0)}\tau_{i,j}^{(2)} \right. \\
&\left. +\tau_{i,j}^{(1)} \tau_{i,j}^{(1)} -2 \tau_{x\,min} (2
\tau_{i,j}^{(1)}+\tau_{i,j}^{(2)})\right) -8 \tau_{i,j}^{(1)} v_{nmo}^2 v_0^2 \sin^2\theta
\left(\Delta x^4 \right.\\
& \times (\tau_{i,j}^{(0)}-\tau_{z\,min})^3-2 \Delta x^2 \Delta z^2 (\tau_{i,j}^{(0)}-\tau_{x\,min}) (\tau_{i,j}^{(0)}-\tau_{z\,min}) \\
& \times \left. \left. \left.  (2
\tau_{i,j}^{(0)}-\tau_{x\,min}-\tau_{z\,min})+\Delta z^4(\tau_{i,j}^{(0)}-\tau_{x\,min})^3\right)\right)\right).
\end{aligned}
$}
\end{equation}

\section{Criterion for causality}
\label{app:B}

In order to ensure causality of the computed traveltime, the update sequence has to be monotone. This amounts to the fact that the updated traveltime value for a particular grid point has to be greater than or equal to the neighboring grid nodes that are used to form the finite-difference stencil. This amounts to the condition~\cite{osher_level_2001}:
\begin{equation}
\frac{\partial \tau}{\partial x} \cdot \frac{\partial H}{\partial p_x} + \frac{\partial \tau}{\partial z} \cdot \frac{\partial H}{\partial p_z} \ge 0,
\label{eq:causality1}
\end{equation}
where $\tau(x,z)$ denotes the traveltime from a source to a receiver point $(x,z)$. $H(x,z,p_x,p_z)$ represents the Hamiltonian, while $p_x,$ and $p_z$ denote the slowness vectors along the $x$, and $z$ directions, respectively.

However, in the case of using one sided finite-difference approximation, as we do in this paper, the criterion given by the inequality~(\ref{eq:causality1}) is equivalent to the stricter Osher's fast marching criterion~\cite{tsai_fast_2003}:
\begin{equation}
\frac{\partial \tau}{\partial x} \cdot \frac{\partial H}{\partial p_x} \ge 0, \qquad \frac{\partial \tau}{\partial z} \cdot \frac{\partial H}{\partial p_z} \ge 0.
\label{eq:causality2}
\end{equation}
This means that the partial derivatives of traveltime $\left(\frac{\partial \tau}{\partial x}, \frac{\partial \tau}{\partial z}\right)$ and their corresponding components of the characteristics directions $\left(\frac{\partial H}{\partial p_x},  \frac{\partial H}{\partial p_z}\right)$ have the same sign. Hence, in order to be accepted as valid solution, the traveltime solution $\tau_{i,j}$ has to satisfy the causality criterion given by~(\ref{eq:causality2}).

\bibliographystyle{model1-num-names}
\bibliography{polynomial_perturbation}

\end{document}